\documentclass[sigconf]{acmart}
\AtBeginDocument{%
  }

\usepackage{changepage}
\usepackage{threeparttable}
\usepackage{multirow}
\usepackage{graphicx}
\usepackage{listings}
\usepackage{xcolor}
\usepackage{enumitem}   
\usepackage{booktabs}   

\newcommand{\quickta}{\textsc{QuickTA}}

\copyrightyear{2026}
\acmYear{2026}
\setcopyright{cc}
\setcctype{by-nc-nd}
\acmConference[ITiCSE 2026]{Proceedings of the 31st ACM Conference on Innovation and Technology in Computer Science Education V. 1}{July 10--15, 2026}{Madrid, Spain}
\acmBooktitle{Proceedings of the 31st ACM Conference on Innovation and Technology in Computer Science Education V. 1 (ITiCSE 2026), July 10--15, 2026, Madrid, Spain}
\acmDOI{10.1145/3803400.3809330}
\acmISBN{979-8-4007-2634-7/2026/07}

\title[Personalizing Computer Science Worksheets with Large Language Models]{Beyond One-Size-Fits-All Exercises: Personalizing Computer Science Worksheets with Large Language Models}

\author{Franco Ortiz}
\authornote{Both authors contributed equally to this work.}
\orcid{0009-0004-4759-5408}
\affiliation{%
  \institution{University of Toronto Mississauga}
  \city{Mississauga}
  \state{Ontario}
  \country{Canada}
}
\email{franco.ortiz@mail.utoronto.ca}

\author{Runlong Ye}
\authornotemark[1]
\orcid{0000-0003-1064-2333}
\affiliation{
\institution{University of Toronto}
   \city{Toronto}
   \state{Ontario}
   \country{Canada}
}
\email{harryye@cs.toronto.edu}

\author{Michael Liut}
\orcid{0000-0003-2965-5302}
\affiliation{%
  \institution{University of Toronto Mississauga}
  \city{Mississauga}
  \state{Ontario}
  \country{Canada}
}
\email{michael.liut@utoronto.ca}

\begin{document}



\begin{abstract}
\textbf{Motivation:}
Large Language Models (LLMs) have been widely applied to student-facing educational tools, this work explores their use in supporting instructors by presenting a practical adaptation of the Framework for Adaptive Content using Educational Technology (FACET) system to generate personalized instructional materials for an Introduction to Computer Programming (CS1) course.

\textbf{Method:}
We conducted a mixed-methods study with $409$ first-year computer science (CS) students, focusing on \textit{regular expressions (RegEx)}. Students were assessed on their knowledge and motivation, classified into one of four learner profiles, and assigned either LLM-personalized (treatment) or standard non-adaptive (control) exercises. Personalized materials varied in scaffolding, instructional explicitness, and tone based on learner profiles grounded in Bloom's Taxonomy and Self-Determination Theory.

\textbf{Results:}
Quantitative analysis reveals that standard exercises resulted in task incompletion among low-knowledge learners, with approximately 25–30\% incompletion, whereas personalized materials sustained near-universal completion ($>99\%$) across all profiles. While high-performing students experienced ceiling effects, Low Knowledge/Low Motivation students achieved significantly higher correctness ($+18.2\%$) with personalized support. Survey data indicate that students prioritize structural scaffolding (logical sequence, difficulty pacing) over motivational tone and perceive the adaptive tasks as equally challenging as standard exercises.

\textbf{Implications:}
These findings suggest that learner–profile–driven LLM personalization primarily serves as a retention scaffold, preventing task abandonment among at-risk students without diminishing the task's ``desirable difficulty''. The results demonstrate that instructor-facing LLM systems can effectively close engagement gaps in CS1 by tailoring instructional explicitness to student needs.

\end{abstract}


\maketitle

\section{Introduction}
The increasing heterogeneity of student populations in introductory computer science (CS1) courses poses a significant pedagogical challenge. Students enter with diverse cognitive abilities and motivational dispositions, yet traditional ``one-size-fits-all'' materials often fail to address these varying needs. While educational psychology emphasizes the importance of differentiating instruction based on cognitive load \cite{duran2022cognitive} and motivational drivers \cite{deci2008self, ryan2020intrinsic}, creating personalized content at scale remains resource-prohibitive for most instructors. Large Language Models (LLMs) offer a potential solution; however, current research predominantly focuses on student-facing tools (e.g., chatbots) rather than instructor-facing systems that rigorously integrate pedagogical frameworks.

This study addresses this gap by presenting a practical operationalization of the Framework for Adaptive Content using Educational Technology (FACET) \cite{2025FACET}. We developed an LLM system to generate personalized laboratory worksheets for a CS1 Regular Expressions module. By classifying 409 students into four profiles based on Bloom’s Taxonomy \cite{AndersonKrathwohl2001} and Self-Determination Theory \cite{deci2008self}, our system adapted task scaffolding, tone, and explicitness to individual learner needs. We conducted a mixed-methods experiment comparing these personalized materials against standard exercises. Our results indicate that while high-performing students succeed regardless of the medium, profile-driven personalization significantly boosts task completion for all low-knowledge learners, and improves correctness specifically for those with low knowledge and low motivation. This work highlights the feasibility of using LLMs not just as content generators, but as pedagogically grounded aids for instructional design.

Our research questions (RQs) are:

\begin{adjustwidth}{2em}{}
\begin{enumerate}
    \item [\textbf{RQ1}:] How does LLM-generated, learner profile-aligned, instructional guidance affect students' engagement and performance compared to non-adaptive exercises? 
    \item [\textbf{RQ2}:] How do students with different knowledge and motivation profiles experience and perceive LLM-personalized exercises? 
\end{enumerate}
\end{adjustwidth}

\section{Related Work}


The increasing heterogeneity of student populations in computer science poses significant challenges for instructors, particularly in foundational CS1 courses where students exhibit diverse cognitive abilities, motivational dispositions, and prior knowledge. While educational theory has long emphasized the importance of personalized, differentiated instruction that addresses both cognitive and affective learner dimensions, translating these principles into scalable classroom practice remains an enduring challenge. 

\subsection{Pedagogically Adaptive Content}

Research in educational psychology has emphasized that effective instruction must account for cognitive and motivational differences among learners. Self-Determination Theory (SDT) distinguishes autonomous and controlled forms of motivation and identifies autonomy, competence, and relatedness as key drivers of sustained engagement and performance in educational contexts \cite{deci2008self, ryan2020intrinsic}.

In computer science education, recent work has highlighted the importance of measuring motivational constructs alongside cognitive outcomes. \citet{lakanen2023cs1} demonstrate that motivation, engagement, and self-efficacy are significant predictors of student success in CS1 courses, while \citet{duran2022cognitive} argue that cognitive load measures should be complemented with motivational variables to better explain learning outcomes.

From an instructional design perspective, Bloom's Taxonomy \cite{AndersonKrathwohl2001} provides a widely adopted framework for aligning task complexity with learner readiness, enabling systematic scaffolding from lower- to higher-order cognitive skills. When combined with motivational considerations, these frameworks support the design of learning experiences that are both cognitively appropriate and supportive of learner engagement.

\subsection{LLM as Educational Support}

The integration of large language models (LLMs) into educational technology has expanded the feasibility of delivering personalized learning at scale~\cite{sharma2025role}. Prior approaches to adaptive instruction, including Intelligent Tutoring Systems (ITS)~\cite{alrakhawi2023intelligent} and Computer-Aided Instruction (CAI)~\cite{suson2020computer}, have demonstrated effectiveness in supporting learners but typically require extensive expert authoring and rule-based design, limiting their scalability.

LLMs introduce new capabilities for educational content generation, personalized feedback, and adaptive instruction. Unlike traditional systems, LLMs can generate contextually appropriate explanations, create novel exercises, and adapt their communication style to individual learners without requiring explicit programming of pedagogical rules \cite{10.1145/3632620.3671103}. These affordances have led to the rapid adoption of LLMs across educational contexts.

Existing applications of LLMs in education can be broadly categorized into student-facing and teacher-facing tools~\cite{2025FACET}. Student-facing systems, such as conversational tutors, homework assistants, and code explanation tools, are now widely used \cite{kazemitabaar2024codeaid, vadaparty2024cs1}. In contrast, comparatively little work has focused on teacher-facing applications that support instructional design and material preparation. As a result, many current systems lack systematic integration of pedagogical frameworks, explicit learner modelling, and quality assurance mechanisms.

The FACET framework~\cite{2025FACET} addresses this gap through a teacher-centred architecture. FACET employs learner agents that represent diverse student profiles across cognitive and motivational dimensions, a teacher agent that generates instructional materials grounded in didactical principles, and an evaluator agent that assesses output quality. This design offers advantages over single-LLM approaches, including improved output stability, reusable learner profiles, and explicit inclusion of pedagogical reasoning.

Recent work in learning analytics further highlights the challenges of deploying adaptive systems in authentic educational settings. \citet{2025MusabirovAdaptiveXPrizekumar2024computer} report lessons learned from large-scale, platform-based adaptive experimentation, emphasizing the complexity of decision-making, the need for instructor-centered tools, and the difficulty of operationalizing personalization while balancing fairness, scalability, and pedagogical intent. Their findings underscore the importance of structured, tool-supported approaches for enabling instructors to design, deploy, and evaluate adaptive interventions in real classrooms.

\subsection{Student Modelling and Learner Profiling}

Effective personalization depends on accurate models of learner characteristics~\cite{chen2024large}. Traditional student modelling has focused primarily on cognitive dimensions, such as knowledge acquisition and skill mastery, often operationalized through knowledge tracing or taxonomy-based stratification of learner readiness~\cite{AndersonKrathwohl2001}.

However, cognitive modelling alone provides an incomplete picture of learner needs. Prior work emphasizes the importance of incorporating affective and motivational dimensions into learner models, as these factors influence engagement and learning behaviours. The Intrinsic Motivation Inventory (IMI)~\cite{ryan1982} offers a validated instrument for measuring intrinsic motivation, enabling more nuanced learner profiles when combined with cognitive assessments.

Recent empirical studies of learner--LLM interaction underscore this limitation. \citet{kumar2024guiding} shows that instructional guidance shapes learners' interaction patterns, confidence, and trust when working with LLM-based tutors, highlighting the role of motivational and metacognitive states beyond cognitive proficiency alone.

The case for multi-dimensional learner modelling is reinforced by links between cognitive load and motivation. As \cite{ryan2020intrinsic} argues, cognitive measures should be accompanied by complementary motivational constructs, an approach adopted in this study. Building on this line of work, we operationalize a two-dimensional learner profile combining knowledge and motivation within an instructor-facing LLM system, inspired by the FACET framework~\cite{2025FACET}.

\section{Methods}

In this section, we detail the design, participants, materials, procedures, and analytical methods of our study. This was conducted in a large undergraduate introductory programming course (CS1) at a research university in North America. CS1 employs an inverted classroom \cite{lage2000inverting} 12-week format, uses Python as the instructional programming language, and the content is aligned with CS2023 curricular recommendations \cite{garcia202520,kumar2024computer}. 

In this CS1 classroom, students completed one in-person term test (20\%), a final exam (35\%), eleven weekly labs (3\% each), and preparatory/practice exercises with videos (12\%) using our in-house learning management system. Completing both the start- and end-of-term surveys earned a 2\% bonus. A total of 1,050 students were enrolled in the course. Most students were first-year undergraduates intending to pursue the computer science major and learned the same content at the same pace.

\subsection{Study Context and Design}

\begin{figure}[htbp]
    \centering
    \includegraphics[width=1\linewidth]{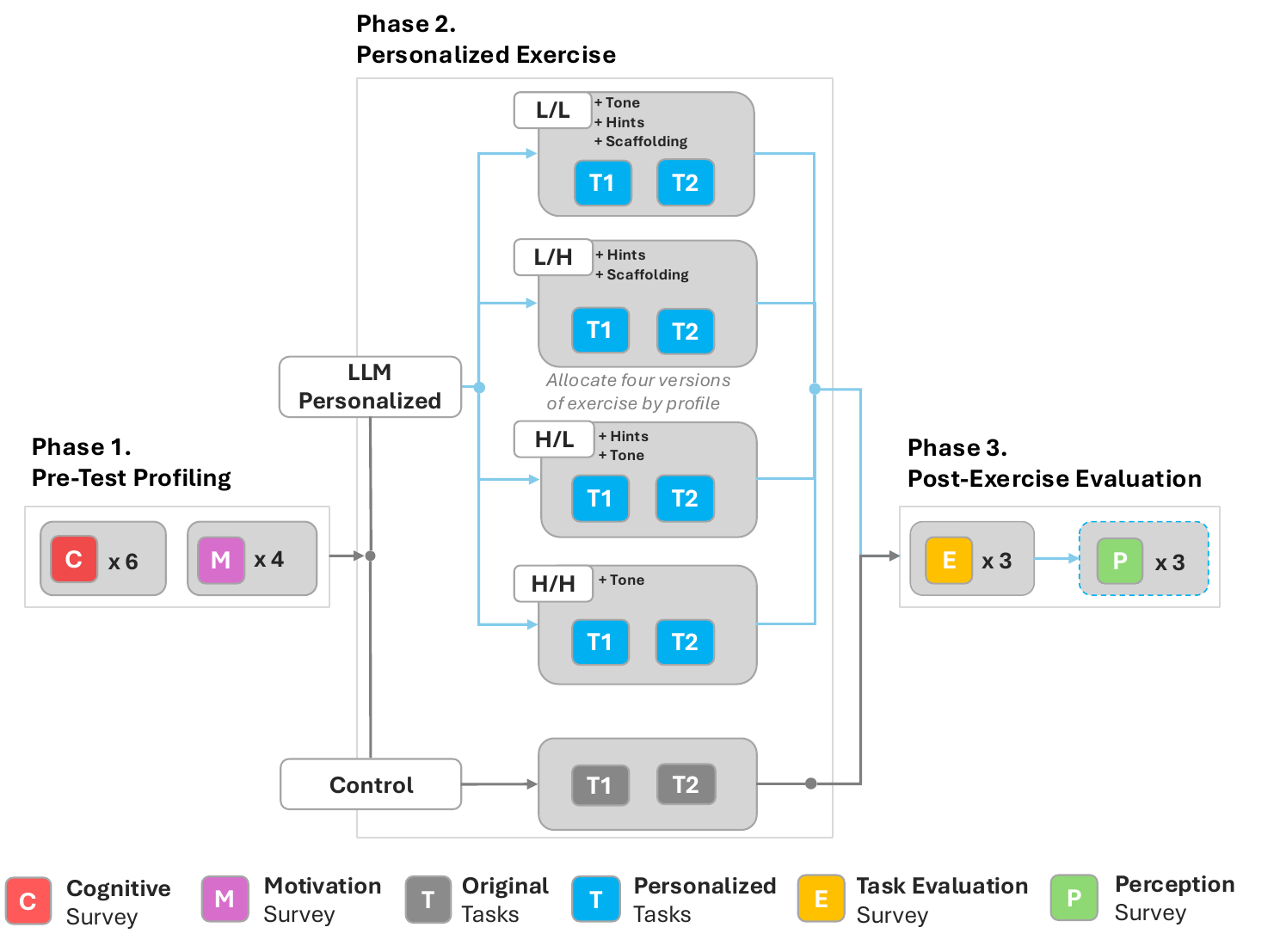}
    \caption{Study workflow: 1. Pre-Test Profiling, 2. Personalized Assignment Allocation, and 3. Post-Exercise Evaluation.}
    \label{fig:stageprocess}
\end{figure}

The study was divided into three phases: pre-test assessment for student profile modelling, personalized task allocation, and post-assignment evaluation. All components were integrated into an educational AI platform (\quickta~\cite{quickta}) that enabled automated profile classification through real-time scoring and threshold-based categorization, dynamic content allocation based on randomization, and immediate delivery of the post-assignment evaluation upon task completion. This integrated technical architecture ensured standardized administration, eliminated manual scoring errors, and preserved experimental integrity.
This stage process is illustrated in~\autoref{fig:stageprocess}.
The complete study materials, including the pre-test assessment, cognitive and motivation items, worksheet examples, and post-exercise evaluation instrument, are available on OSF\footnote{\url{https://osf.io/bmv48/}}.
\subsection{Phase 1: Pre-Test Profiling}

To ensure each student received instructional materials tailored to their individual profile, we conducted a comprehensive pre-test measuring two dimensions: cognitive ability \& intrinsic motivation.

\subsubsection{Cognitive Assessment}

The cognitive assessment consisted of six questions designed following Bloom's taxonomy \cite{AndersonKrathwohl2001}, progressing from general computer science knowledge to specific regular expression concepts. This stratification allowed us to directly evaluate students' cognitive capacity across various levels of complexity. Questions were weighted differentially based on cognitive demand: foundational questions (Questions 1--2) were worth 1 point each, application questions (Questions 3--4) were worth 2 points each, an analysis question (Question 5) was worth 3 points, and a synthesis question (Question 6) was worth 4 points, yielding a maximum possible score of 13 points. 

\subsubsection{Motivation Assessment}

Intrinsic motivation was measured using a 4-item questionnaire adapted from the Intrinsic Motivation Inventory (IMI) \cite{ryan2020intrinsic}. Students responded to items reflecting interest/enjoyment, perceived competence, perceived choice, and autonomy using a 5-point Likert scale. Each item was weighted equally at 25\%, and responses were aggregated to produce an average motivation score ranging from 1 to 5.

Given the survey-length constraints of an in-class deployment, we selected one item from three relevant IMI sub-scales (out of seven) to obtain a broad motivational snapshot while minimizing respondent burden: \textit{Interest/Enjoyment} (1 Question), \textit{Perceived Competence} (1 Question), and \textit{Pressure/Tension} (2 Questions)\cite{ryan1982}. Items were selected to tap distinct motivational dimensions relevant to the RegEx task context.
Because single-item sub-scale scores preclude standard reliability estimation, we treat responses at the item level and aggregate them into a composite motivation score.

\subsection{Profile Classification}
\label{profileclass}

Students were categorized into \textit{four distinct learner profiles} based on threshold-based classification: $\geq$10/13 points (77\%) for knowledge and $\geq$3/5 points (60\%) for motivation; defined as follows:
\begin{adjustwidth}{2em}{}
\begin{enumerate}
    \item[] \textbf{L/L}: Low Knowledge/Low Motivation
    \item[] \textbf{L/H}: Low Knowledge/High Motivation
    \item[] \textbf{H/L}: High Knowledge/Low Motivation
    \item[] \textbf{H/H}: High Knowledge/High Motivation
\end{enumerate}
\end{adjustwidth}

\subsubsection{Knowledge Classification}


Students scoring $\geq$10/13 points (77\%) were classified as high knowledge, 
consistent with mastery learning benchmarks commonly used in CS education 
\cite{asee2004, nebraska2020}, while those below this threshold were classified 
as low knowledge.

\subsubsection{Motivation Classification}


Students scoring $\geq$3/5 points (60\%) were classified as high motivation. This midpoint-plus-margin classification distinguishes students whose intrinsic motivation sustains engagement through demanding tasks from those requiring additional support, consistent with prior CS1 research linking higher intrinsic motivation to better academic outcomes \cite{Rutherford24112024}.

\subsubsection{Final Sample Distribution by Profile}

\autoref{tab:descriptives} presents the descriptive statistics for each learner profile, demonstrating successful differentiation along both dimensions.

\vspace{-0.25cm}

\begin{table}[h]
\centering
\caption{Descriptive Statistics by Learner Profile ($N = 409$)}
\label{tab:descriptives}
\begin{tabular}{|c||c|c|c|}
\hline
\multirow{2}{*}{\textbf{Profile}} & \multirow{2}{*}{\textbf{$N$ (\%)}} & \textbf{Knowledge} & \textbf{Motivation} \\
 & & \textbf{$M$ ($SD$)} & \textbf{$M$ ($SD$)} \\
\hline\hline
L/L & 107 (26.2\%) & 6.11 (2.50) & 2.69 (0.64) \\
L/H & 72 (17.6\%) & 6.46 (2.13) & 3.62 (0.48) \\
H/L & 118 (28.9\%) & 11.44 (1.22) & 2.83 (0.59) \\
H/H & 112 (27.4\%) & 11.50 (1.38) & 3.70 (0.62) \\
\hline
\end{tabular}
\begin{tablenotes}
\small
\item \textit{Note.} $M$ = Mean; $SD$ = Standard Deviation. Learner Profile formatted in Knowledge-level/Motivation-level (see~\autoref{profileclass}). The number of students is acquired after filtering for attention checks.
\end{tablenotes}
\end{table}

\vspace{-0.5cm}

\subsection{Phase 2: Personalized Exercise}

\subsubsection{Randomized Assignment}

After profile classification, students were randomly assigned (approximately 50/50 split) to one of two conditions: students in the LLM-personalized condition received exercises adapted to their specific learner profile, while students in the control condition received neutral, non-adapted baseline exercises. Students were blind to their assigned condition and were not informed whether they received LLM-generated or standard materials, reducing potential placebo effects and response bias. The core learning objectives and fundamental task structure remained constant across all versions to ensure valid comparison.

\subsubsection{Adaptation Framework}

The personalized exercises were adapted along three instructional dimensions, inspired by FACET \cite{2025FACET}:
\begin{enumerate}
    \item \textbf{Scaffolding level:} High scaffolding (progressive hints, worked examples of input and/or output) for low-knowledge learners; minimal scaffolding for high-knowledge learners
    
    \item \textbf{Instructional tone:} Motivational and encouraging tone with real-world contexts for low-motivation students; neutral-professional tone for high-motivation students
    
    \item \textbf{Instruction explicitness:} Explicit, detailed step-by-step guidance for low-knowledge students; implicit, challenge-oriented instructions for high-knowledge students
\end{enumerate}

Note that the control worksheet contained two regular expression (RegEx) tasks without any instructional adaptations. 

\subsubsection{Content Generation via LLM-Based System}

Personalized worksheets were generated using Claude Sonnet 4.5 \cite{anthropic2025claudesonnet45} through a carefully engineered prompt designed to implement the adaptation framework systematically. The prompt incorporated student profile information (knowledge and motivation levels), Bloom's taxonomy alignment for task difficulty selection, and explicit instructions for personalization along the three adaptation dimensions.

The system prompt instructed the LLM to present a clear assessment of the student's learning situation based on pre-test results, and to adapt tasks aligned with Bloom's taxonomy, selecting scaffolding and instructional elements appropriate to the student's knowledge level, and providing hints and tone appropriate to the student's motivation level. 

\subsubsection{Material Selection}
This study focused specifically on regular expression instruction, a fundamental yet challenging topic in CS1 curricula \cite{michael2019regexes}. The LLM personalized the content based on students' knowledge and motivation profiles. From the LLM-generated materials, four worksheets were selected, one for each learner profile, through expert review by CS1 instructors to ensure pedagogical quality, accurate regular expression content, and appropriate difficulty calibration. This research was approved by the University of Toronto's Research Ethics Protocols \#38966 and \#43701.


\subsubsection{Sample Characteristics}
The final sample consisted of 409 CS1 students, with 233 students (57.0\%) in the LLM-personalized condition and 176 students (43.0\%) in the control condition. 
High-knowledge students scored around 11 points on the pre-test, while low-knowledge students scored around 6 points. High motivation students scored around 3.6-3.7 on the motivation scale, while low motivation students scored around 2.6-2.9. This distribution provided sufficient cell sizes for comparisons between the LLM-personalized and control groups.

\subsection{Phase 3: Post-Exercise Evaluation}

All students completed a post-exercise survey with three common items assessing perceived difficulty, understanding of regular expressions, and usefulness of instructional elements (worked examples, hints, real-world context, difficulty progression, and instruction clarity). Students in the LLM-personalized condition received five additional items evaluating their perceptions of personalization, including overall personalization quality, specific aspects noticed, cognitive adaptation, and motivational impact.

\section{Results}
\subsection{Completion Rates}
Prior to data cleaning and filtering, we analyzed initial task completion rates to assess participant engagement differences between conditions. Group sizes were unequal 
(control: 289, LLM-personalized: 234) as a natural outcome of random assignment, 
not by design. Students who abandoned the study did so before starting the RegEx tasks, with no incompletion occurring during task completion or the post-survey phase.

81.3\% (235 out of 289 students) of students from the control group completed the exercise, whereas 99.5\% (233 out of 234 students) of the LLM-personalized group completed the exercise (including those who failed attention checks).

\begin{figure}[htbp]
    \centering
    \includegraphics[width=.85\linewidth]{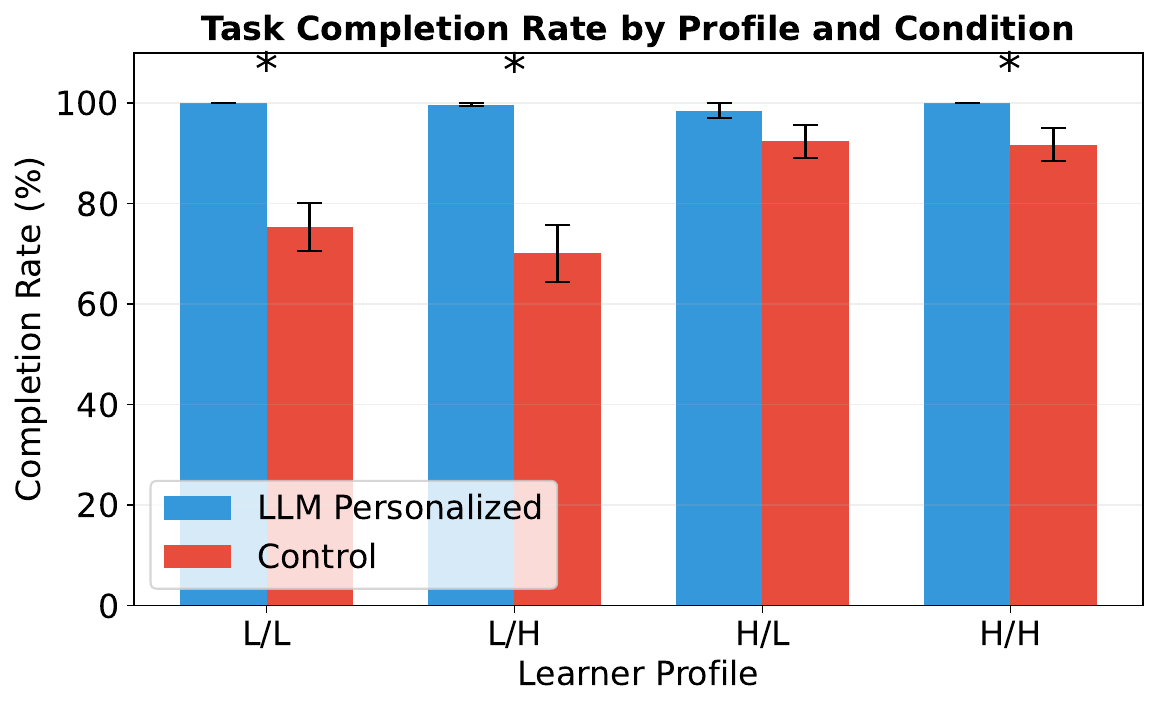}
    \caption{Task Completion by Learner Profile and Experiment Condition. Error bars represent standard error. *Significant difference at p < .05. Learner Profile formatted in Knowledge-level/Motivation-level (see~\autoref{profileclass}).}
    \label{fig:placeholder}
\end{figure}

A notable contrast in engagement emerged between the two conditions. While the LLM-personalized group achieved near-perfect completion rates, the control group exhibited significant task incompletion that varied by learner profile ($\chi^2(3) = 18.38, p < .001$). In \autoref{fig:placeholder}, the personalized intervention effectively closed the engagement gap: LLM participants maintained completion rates between 98.5\% and 100\% regardless of their profile. Conversely, control participants with low knowledge struggled significantly, with completion plummeting to 75.3\% for the L/L profile and 70.1\% for the L/H profile, roughly 20 percentage points lower than their high-knowledge peers (H/H: 91.7\%, H/L: 92.4\%) and nearly 30 points lower than their counterparts in the LLM condition.


\subsection{Performance Evaluation}
To assess performance, we evaluated student responses using functional matching tests for each task. A response was marked as correct if the submitted answer produced the expected output. To account for equivalent but syntactically different solutions, we implemented a robust multi-tiered evaluation framework that recognizes valid variations of the correct solution.

Within the control group, performance varied significantly across profiles (Kruskal-Wallis H=8.229, p=.042), ranging from 54.4\% (L/L) to 71.9\% (H/H). In contrast, the LLM-personalized group showed no significant performance differences across profiles (H=4.242, p=.237), with correctness ranging from 60.8\% (H/L) to 72.6\% (L/L), as shown in \autoref{fig:correctness}.

Direct comparisons between LLM-personalized and control conditions within each profile revealed that Low Knowledge/Low Motivation students performed significantly better with personalized tasks (72.6\%) compared to standard (Control) tasks (54.4\%; U=1807.0, p=.006). No significant differences emerged for H/H (p=.413), H/L (p=.397), or L/H profiles (p=.491). \autoref{fig:correctness} illustrates these performance differences across learner profiles and conditions.

\begin{figure}[ht]
\centering
\includegraphics[width=.85\linewidth]{correctness_grouped_bar.pdf}
\caption{Student Performance by Learner Profile and Experiment Condition. Error bars represent standard error. *Significant difference at p < .05. Learner Profile formatted in Knowledge-level/Motivation-level (see~\autoref{profileclass}).}
\label{fig:correctness}
\end{figure}

\subsection{Perceived Difficulty and Understanding}
Students in both conditions reported similar perceptions of task difficulty and understanding of regular expressions in the post-survey. Perceived difficulty scores were comparable between the LLM-personalized group (M = 3.89, Mdn = 4.0) and the control group (M = 3.97, Mdn = 4.00; U = 19420.0, p = .389). Self-reported understanding of regular expressions showed no significant difference across conditions (LLM-personalized: M = 3.17, Mdn = 3.0; Control: M = 3.29, Mdn = 3.00; U = 18946.5, p = .199).

Students also rated the usefulness of five learning elements on a 5-point scale (1 = Very useful, 5 = Not present). Responses were similar across conditions for all elements, including step-by-step worked examples, hints/clues in the tasks, real-world context/applications, gradual progression of difficulty, and clear instructions (all p > .05).

\subsection{Perceptions of Personalization (LLM-Personalized Group Only)}
Students in the LLM-personalized group rated their perceptions of personalization, cognitive adaptation, and motivational effects on 6-point scale. While all three measures showed statistically significant differences across learner profiles (perception of personalization: H statistic = 9.32, p = .025; cognitive adaptation: H statistic = 13.03, p = .005; motivation effect: H statistic = 10.11, p = .018), effect sizes were small ($\eta^2$ = 0.040–0.056). As shown in \autoref{tab:motivation_comparison}, high motivation students (L/H, H/H) reported significantly higher scores than low motivation students (L/L, H/L) across all three measures.



\begin{table}[ht]
\centering
\small
\caption{Personalization Perceptions by Motivation Level}
\label{tab:motivation_comparison}
\begin{tabular}{|l||c|c|c|}
\hline
\multirow{2}{*}{\textbf{Measure}} & Mean, \textbf{Low} & Mean, \textbf{High} & \multirow{2}{*}{\textbf{p-value}} \\
& \textbf{Motivation} & \textbf{Motivation} & \\
\hline\hline
Motivation effect & 3.37 & 3.71 & .003* \\
Cognitive adaptation & 3.32 & 3.80 & .001* \\
Perception of personalization & 3.31 & 3.71 & .003* \\
\hline
\multicolumn{4}{l}{\footnotesize *Significant at p < .05} \\
\end{tabular}
\end{table}

Students identified aspects of personalization they noticed through a multiple-select question. ``Difficulty level appropriate'' was most frequently noticed (51.7\%), followed by ``Logical sequence of tasks'' (47.9\%) and ``Examples relevant to interests'' (42.3\%). ``Adequate scaffolding/support'' (39.7\%) and ``Motivational tone'' (26.4\%) were the least frequently identified aspects.

Student responses to personalized exercises were positive, with appreciation for key design elements. Students valued the learner-controlled hint system that promoted active learning rather than providing direct answers, and the structured progression of difficulty that helped build understanding. Students found the exercises engaging and effective for developing RegEx skills, with comments including ``I can finally form RegEx confidently'' and ``this lab helped me exercise my skills and allowed me to see the areas of RegEx I need to improve in.'' However, feedback revealed desires for immediate correctness feedback and a clearer indication of task length.

\section{Discussion}

\subsection{RQ1: Reducing Task Incompletion and Supporting At-Risk Learners}
The most immediate impact of the personalized intervention was on task completion. In the control condition, students with low prior knowledge had high incompletion rates, with completion falling to 75.3\% for Low-Knowledge/Low-Motivation (L/L) students and 70.1\% for Low-Knowledge/High-Motivation (L/H) students. This aligns with prior CS education research showing that initial friction often causes novice programmers to disengage \cite{kinnunen2006students, beaubouef2005high}. In contrast, the LLM-personalized condition achieved near-total completion ($\approx 99\%$) across all profiles. By adjusting the instructional explicitness and tone, the system appeared to lower the effective barrier to entry. This scaffolding likely satisfied the students' psychological need for competence, a key aspect of engagement in Self-Determination Theory \cite{deci2008self}, allowing them to persist through the task where they otherwise would have quit. This aligns with recent findings by \citet{abolnejadian2024leveraging}, who demonstrated 
that LLM-powered adaptive instruction in CS1 contexts improved student engagement, suggesting that personalized scaffolding addresses structural gaps in one-size-fits-all 
instruction.

Beyond retention, the personalization significantly improved performance for the most vulnerable demographic. The L/L students in the personalized group achieved a correctness rate of 72.6\%, compared to just 54.4\% in the control group, an 18.2 percentage point increase. It is important to note that the control group's performance data likely overestimates their true proficiency due to ``survivor bias'': the weakest control students did not complete it, meaning we are comparing the entire personalized cohort against only the most persistent control students.

However, the results for the L/H (Low Knowledge/High Motivation) profile highlight a distinction between \textit{behavioural completion} and \textit{cognitive proficiency}. While the system ensured these students finished the work (99.7\%), their final scores were not significantly higher than the control group. This suggests that while personalization can provide the motivational support to keep students on task, resolving deep knowledge gaps may require iterative practice beyond a single assignment.
This distinction echoes findings from CS1 contexts where LLM assistance improved task completion but did not always translate into independent mastery of programming concepts \cite{10.1145/3649217.3653584}, underscoring the need for iterative rather than single-intervention designs.

\subsection{RQ2: Preference for Structural Support and Desirable Difficulty}
Student feedback suggests that learners adopt a utilitarian mindset when working with AI tools. Although the system explicitly adapted the ``motivational tone'' (e.g., encouraging vs. professional), students were far more likely to notice structural adaptations. Survey results show that students identified the ``Logical sequence of tasks'' (47.9\%) and ``Difficulty level'' (51.7\%) as key features, while only 26.4\% noticed the motivational tone. This aligns with Cognitive Load Theory \cite{sweller1988cognitive}: when novices struggle with complex material such as regular expressions, they prioritize instructional clarity that reduces extraneous load rather than emotional support.

Crucially, the personalized exercises were not perceived as ``easier.'' Students in both groups reported nearly identical levels of perceived difficulty ($M=3.89$ for Personalized vs. $M=3.97$ for Control). This indicates that the system maintained ``desirable difficulty'' \cite{bjork1994memory} and provided enough support to make the task achievable without removing the challenge necessary for learning.

Finally, we observed a metacognitive gap in how personalization was experienced. High-motivation students were significantly more likely to notice the specific adaptations than their low-motivation peers. This creates a paradox: the students who benefited most from the scaffolding (L/L) were the least likely to recognize it. It appears that struggling learners were so consumed by the cognitive demands of the problem-solving process that they lacked the bandwidth to reflect on the instructional design. They utilized the support to survive the task, while high-performing students had the surplus cognitive capacity to analyze the tool itself.
This finding is consistent with \citet{kumar2024guiding} and \citet{ramirez2025understanding, ramirez2026investigating} showing that struggling learners tend to interact with AI tools instrumentally rather than reflectively, prioritizing task completion over engagement with the adaptive features themselves.

\subsection{Limitations and Future Work}

Several limitations warrant consideration: measurement validity and experimental design. Due to strict in-class time constraints, we used an ad hoc cognitive assessment and a shortened four-item adaptation of Intrinsic Motivation, without formal psychometric validation. Consequently, our evaluation relies heavily on self-reported perceptions, making the findings susceptible to Hawthorne effects and response bias. Furthermore, to provide appropriate scaffolding, the number of sub-tasks was deliberately varied across profiles in the personalized condition. While pedagogically justified, this introduces an experimental confound between the type of personalization and the volume of practice. Future work should employ fully validated psychometric scales and complement self-reports with objective behavioural data (e.g., time-on-task, help-seeking patterns) to strengthen construct validity.

Second, this intervention was limited to a single, high-friction topic (RegEx). While this context highlighted attrition patterns, we cannot assume these retention and performance gains will generalize uniformly across the CS1 curriculum. Future longitudinal research should evaluate LLM-driven personalization across multiple foundational concepts to determine if short-term engagement translates into sustained learning gains. Finally, our methodology relied on expert human-in-the-loop selection of the LLM-generated worksheets to ensure pedagogical quality. Transitioning to a fully automated deployment pipeline, potentially leveraging evaluator agents for rigorous quality assurance, remains a critical next step for scaling this instructor-facing approach.

Additionally, a promising direction is integrating small language models (SLMs) to deliver adaptive feedback in the moment of struggle, detect misconceptions as students work, and extend the present retention scaffold into a tighter feedback loop that may help convert sustained engagement into long-lasting learning gains.

\section{Conclusion}

This study offers initial evidence that LLM-generated personalization can effectively mitigate student disengagement in CS1. While standard ``one-size-fits-all'' exercises led to high task incompletion among low-knowledge learners, profile-driven personalization served as a critical retention scaffold. It sustained near-perfect completion rates and significantly improved short-term performance for the most vulnerable students (Low Knowledge/Low Motivation). Importantly, students viewed these adapted materials as structural supports rather than shortcuts, indicating that targeted scaffolding lowers entry barriers while preserving the ``desirable difficulty'' necessary for learning. This work highlights the potential of instructor-facing AI to democratize differentiated instruction, providing educators with a personalizable and scalable tool to close the engagement gap and promote persistence in early computer science education.
\newpage
\begin{acks}
We acknowledge the financial support of the Natural Sciences \& Engineering Research Council of Canada (NSERC), Discovery Grant \#RGPIN-2024-04348 and the University of Toronto's Learning \& Education Advancement Fund (LEAF).
\end{acks}
\balance
\bibliographystyle{ACM-Reference-Format}
\bibliography{references}

@misc{2025FACET,
      title={FACET: Teacher-Centred LLM-Based Multi-Agent Systems-Towards Personalized Educational Worksheets}, 
      author={Jana Gonnermann-Müller and Jennifer Haase and Konstantin Fackeldey and Sebastian Pokutta},
      year={2025},
      eprint={2508.11401},
      archivePrefix={arXiv},
      primaryClass={cs.HC}
}

@article{deci2008self,
  title={Self-determination theory: A macrotheory of human motivation, development, and health.},
  author={Deci, Edward L and Ryan, Richard M},
  journal={Canadian psychology/Psychologie canadienne},
  volume={49},
  number={3},
  pages={182},
  year={2008},
  publisher={Educational Publishing Foundation}
}

@article{ryan2020intrinsic,
  title={Intrinsic and extrinsic motivation from a self-determination theory perspective: Definitions, theory, practices, and future directions},
  author={Ryan, Richard M and Deci, Edward L},
  journal={Contemporary educational psychology},
  volume={61},
  pages={101860},
  year={2020},
  publisher={Elsevier}
}

@article{lakanen2023cs1,
  title={CS1: Intrinsic motivation, self-efficacy, and effort},
  author={Lakanen, Antti-Jussi and Isom{\"o}tt{\"o}nen, Ville},
  journal={Informatics in Education},
  volume={22},
  number={4},
  pages={651--670},
  year={2023},
  publisher={Vilnius University Institute of Data Science and Digital Technologies}
}

@article{duran2022cognitive,
  title={Cognitive load theory in computing education research: A review},
  author={Duran, Rodrigo and Zavgorodniaia, Albina and Sorva, Juha},
  journal={ACM Transactions on Computing Education (TOCE)},
  volume={22},
  number={4},
  pages={1--27},
  year={2022},
  publisher={ACM New York, NY}
}

@article{Rutherford24112024,
  title={How self-beliefs, values, and belonging change and relate with performance during introductory computer science},
  author={Rutherford, Teomara and Lee, Hye Rin and Bart, Austin Cory and Rodrigues, Andrew and Englert, Megan},
  journal={Computer Science Education},
  volume={36},
  number={1},
  pages={166--202},
  year={2026},
  publisher={Taylor \& Francis}
}

@inproceedings{10.1145/3649217.3653584,
author = {Vadaparty, Annapurna and Zingaro, Daniel and Smith IV, David H. and Padala, Mounika and Alvarado, Christine and Gorson Benario, Jamie and Porter, Leo},
title = {CS1-LLM: Integrating LLMs into CS1 Instruction},
year = {2024},
isbn = {9798400706004},
pages = {297–303},
numpages = {7},
location = {Milan, Italy},
series = {ITiCSE 2024}
}

@book{AndersonKrathwohl2001,
  title     = {A Taxonomy for Learning, Teaching, and Assessing: A Revision of Bloom's Taxonomy of Educational Objectives},
  editor    = {Anderson, Lorin W. and Krathwohl, David R.},
  year      = {2001},
  publisher = {Longman},
  address   = {New York},
  edition   = {Complete Edition}
}

@article{ryan1982,
    author = {Ryan, Richard M.},
    title = {Control and Information in the Intrapersonal Sphere: An Extension of Cognitive Evaluation Theory},
    journal = {Journal of Personality and Social Psychology},
    volume = {43},
    number = {3},
    pages = {450--461},
    year = {1982}
}

@inproceedings{asee2004,
  title={CS1 and CS2 Programming Exams for Assessing Learning and Teaching},
  author={Stockman, George and Albee, Paul and Dillon, Laura and Oleszkiewicz, Jonathon},
  booktitle={2004 Annual Conference},
  pages={9--358},
  year={2004}
}

@misc{nebraska2020,
    author = {{University of Nebraska Center for Transformative Teaching}},
    title = {What is Mastery Grading?},
    year = {2020},
    url = {https://teaching.unl.edu/resources/alternative-grading/mastery-grading/},
    note = {Accessed: 2025-01-16}
}

@article{lage2000inverting,
  title={Inverting the classroom: A gateway to creating an inclusive learning environment},
  author={Lage, Maureen J and Platt, Glenn J and Treglia, Michael},
  journal={The journal of economic education},
  volume={31},
  number={1},
  pages={30--43},
  year={2000},
  publisher={Taylor \& Francis}
}

@misc{kumar2024computer,
  title={Computer science curricula 2023},
  author={Kumar, Amruth N and Raj, Rajendra K and Aly, Sherif G and Anderson, Monica D and Becker, Brett A and Blumenthal, Richard L and Eaton, Eric and Epstein, Susan L and Goldweber, Michael and Jalote, Pankaj and others},
  year={2024},
  publisher={ACM}
}

@article{garcia202520,
  title={20 Years Later: A Replication Study on Teaching CS1 Concepts},
  author={Garcia, Rita and Craig, Michelle},
  journal={ACM Transactions on Computing Education},
  volume={25},
  number={2},
  pages={1--33},
  year={2025},
  publisher={ACM New York, NY}
}

@misc{anthropic2025claudesonnet45,
  title = {Introducing Claude Sonnet 4.5},
  author = {Anthropic},
  year = {2025},
  month = sep,
  howpublished = {\url{https://www.anthropic.com/news/claude-sonnet-4-5}},
  note = {Accessed: 2026-01-19}
}

@inproceedings{michael2019regexes,
  title={Regexes are hard: Decision-making, difficulties, and risks in programming regular expressions},
  author={Michael, Louis G and Donohue, James and Davis, James C and Lee, Dongyoon and Servant, Francisco},
  booktitle={2019 34th IEEE/ACM International Conference on Automated Software Engineering (ASE)},
  pages={415--426},
  year={2019},
  organization={IEEE}
}

@article{sharma2025role,
  title={The role of large language models in personalized learning: a systematic review of educational impact},
  author={Sharma, Sahil and Mittal, Puneet and Kumar, Mukesh and Bhardwaj, Vivek},
  journal={Discover Sustainability},
  volume={6},
  number={1},
  pages={1--24},
  year={2025},
  publisher={Springer}
}

@article{alrakhawi2023intelligent,
  title={Intelligent tutoring systems in education: a systematic review of usage, tools, effects and evaluation},
  author={Alrakhawi, Hazem A and Jamiat, Nurullizam and Abu-Naser, Samy S},
  journal={Journal of Theoretical and Applied Information Technology},
  volume={101},
  number={4},
  pages={1205--1226},
  year={2023}
}

@inproceedings{2025MusabirovAdaptiveXPrizekumar2024computer,
author = {Musabirov, Ilya and Reza, Mohi and Song, Haochen and Moore, Steven and Chen, Pan and Kumar, Harsh and Li, Tong and Stamper, John and Bier, Norman and Rafferty, Anna and Price, Thomas and Deliu, Nina and Durand, Audrey and Liut, Michael and Williams, Joseph Jay},
title = {Platform-based Adaptive Experimental Research in Education: Lessons Learned from The Digital Learning Challenge},
year = {2025},
isbn = {9798400707018},
url = {https://doi.org/10.1145/3706468.3706471},
doi = {10.1145/3706468.3706471},
booktitle = {Proceedings of the 15th International Learning Analytics and Knowledge Conference},
pages = {13–23},
numpages = {11},
location = {
},
series = {LAK '25}
}

@article{suson2020computer,
  title={Computer aided instruction to teach concepts in education},
  author={Suson, Roberto and Ermac, Eugenio A},
  journal={International Journal on Emerging Technologies},
  year={2020}
}

@article{chen2024large,
  title={When large language models meet personalization: Perspectives of challenges and opportunities},
  author={Chen, Jin and Liu, Zheng and Huang, Xu and Wu, Chenwang and Liu, Qi and Jiang, Gangwei and Pu, Yuanhao and Lei, Yuxuan and Chen, Xiaolong and Wang, Xingmei and others},
  journal={World Wide Web},
  volume={27},
  number={4},
  pages={42},
  year={2024},
  publisher={Springer}
}

@article{kumar2024guiding,
  title={Guiding Students in Using LLMs in Supported Learning Environments: Effects on Interaction Dynamics, Learner Performance, Confidence, and Trust},
  author={Kumar, Harsh and Musabirov, Ilya and Reza, Mohi and Shi, Jiakai and Wang, Xinyuan and Williams, Joseph Jay and Kuzminykh, Anastasia and Liut, Michael},
  journal={Proceedings of the ACM on Human-Computer Interaction},
  volume={8},
  number={CSCW2},
  pages={1--30},
  year={2024},
  publisher={ACM New York, NY, USA}
}

@inproceedings{10.1145/3632620.3671103,
author = {Logacheva, Evanfiya and Hellas, Arto and Prather, James and Sarsa, Sami and Leinonen, Juho},
title = {Evaluating Contextually Personalized Programming Exercises Created with Generative AI},
year = {2024},
isbn = {9798400704758},
booktitle = {Proceedings of the 2024 ACM Conference on International Computing Education Research - Volume 1},
pages = {95–113},
numpages = {19},
location = {Melbourne, VIC, Australia},
series = {ICER '24}
}

@article{beaubouef2005high,
  title={Why the high attrition rate for computer science students: some thoughts and observations},
  author={Beaubouef, Theresa and Mason, John},
  journal={ACM SIGCSE Bulletin},
  volume={37},
  number={2},
  pages={103--106},
  year={2005},
  publisher={ACM New York, NY, USA}
}

@inproceedings{kinnunen2006students,
  title={Why students drop out CS1 course?},
  author={Kinnunen, P{\"a}ivi and Malmi, Lauri},
  booktitle={Proceedings of the second international workshop on Computing education research},
  pages={97--108},
  year={2006}
}

@article{sweller1988cognitive,
  title={Cognitive load during problem solving: Effects on learning},
  author={Sweller, John},
  journal={Cognitive science},
  volume={12},
  number={2},
  pages={257--285},
  year={1988},
  publisher={Elsevier}
}

@article{bjork1994memory,
  title={Memory and metamemory considerations in the training of human beings},
  author={Bjork, Robert A},
  journal={Metacognition: Knowing about knowing},
  volume={185},
  number={7.2},
  pages={185--205},
  year={1994}
}

@inproceedings{quickta,
  title={Quickta: exploring the design space of using large language models to provide support to students},
  author={Kumar, Harsh and Musabirov, Ilya and Williams, Joseph Jay and Liut, Michael},
  year={2023},
  organization={Learning Analytics and Knowledge Conference 2023 (LAK’23)}
}

@inproceedings{abolnejadian2024leveraging,
  title={Leveraging ChatGPT for adaptive learning through personalized prompt-based instruction: A CS1 education case study},
  author={Abolnejadian, Mohammad and Alipour, Sharareh and Taeb, Kamyar},
  booktitle={Extended abstracts of the CHI conference on human factors in computing systems},
  pages={1--8},
  year={2024}
  }

@inproceedings{kazemitabaar2024codeaid,
  title={Codeaid: Evaluating a classroom deployment of an llm-based programming assistant that balances student and educator needs},
  author={Kazemitabaar, Majeed and Ye, Runlong and Wang, Xiaoning and Henley, Austin Zachary and Denny, Paul and Craig, Michelle and Grossman, Tovi},
  booktitle={Proceedings of the 2024 chi conference on human factors in computing systems},
  pages={1--20},
  year={2024}
}

@incollection{vadaparty2024cs1,
  title={Cs1-llm: Integrating llms into cs1 instruction},
  author={Vadaparty, Annapurna and Zingaro, Daniel and Smith IV, David H and Padala, Mounika and Alvarado, Christine and Gorson Benario, Jamie and Porter, Leo},
  booktitle={Proceedings of the 2024 on Innovation and Technology in Computer Science Education v. 1},
  pages={297--303},
  year={2024}
}

@inproceedings{ramirez2025understanding,
  title={Understanding the Impact of Using Generative AI Tools in a Database Course},
  author={Ramirez Osorio, Valeria and Zavaleta Bernuy, Angela and Simion, Bogdan and Liut, Michael},
  booktitle={Proceedings of the 56th ACM Technical Symposium on Computer Science Education V. 1},
  pages={959--965},
  year={2025}
}

@inproceedings{ramirez2026investigating,
  title={Investigating the Impact of Student Usage of Generative AI Tools in Computing Courses},
  author={Ramirez Osorio, Valeria and Ben Haim, Ido and Ashraf, Ahmed and Mahmoud, Mohammad and Dixon, Peter and Simion, Bogdan and Liut, Michael and Zavaleta Bernuy, Angela},
  booktitle={Proceedings of the 31st ACM Conference on Innovation and Technology in Computer Science Education V. 1},
  year={2026}
}


\end{document}